\documentstyle[twoside,epsf,psfig,times]{article}
\def\beqa{\begin{eqnarray}}
\def\eeqa{\end{eqnarray}}
\def\beq{\begin{equation}}
\def\eeq{\end{equation}}
\catcode`\@=11
\long\def\@makefntext#1{
\protect\noindent \hbox to 3.2pt {\hskip-.9pt  
$^{{\eightrm\@thefnmark}}$\hfil}#1\hfill}               

\def\@makefnmark{\hbox to 0pt{$^{\@thefnmark}$\hss}}    

\def\ps@myheadings{\let\@mkboth\@gobbletwo
\def\@oddhead{\hbox{}
\rightmark\hfil\eightrm\thepage}   
\def\@oddfoot{}\def\@evenhead{\eightrm\thepage\hfil
\leftmark\hbox{}}\def\@evenfoot{}
\def\sectionmark##1{}\def\subsectionmark##1{}}

\oddsidemargin=\evensidemargin
\newcounter{sectionc}\newcounter{subsectionc}\newcounter{subsubsectionc}
\renewcommand{\section}[1] {\vspace{12pt}\addtocounter{sectionc}{1} 
\setcounter{subsectionc}{0}\setcounter{subsubsectionc}{0}\noindent 
        {\tenbf\thesectionc. #1}\par\vspace{5pt}}
\renewcommand{\subsection}[1] {\vspace{12pt}\addtocounter{subsectionc}{1} 
\setcounter{subsubsectionc}{0}\noindent 
{\bf\thesectionc.\thesubsectionc. {\kern1pt \bfit #1}}\par\vspace{5pt}}
\renewcommand{\subsubsection}[1] {\vspace{12pt}\addtocounter{subsubsectionc}{1}
        \noindent{\tenrm\thesectionc.\thesubsectionc.\thesubsubsectionc.
        {\kern1pt \tenit #1}}\par\vspace{5pt}}
\newcommand{\nonumsection}[1] {\vspace{12pt}\noindent{\tenbf #1}
        \par\vspace{5pt}}

\newcounter{appendixc}
\newcounter{subappendixc}[appendixc]
\newcounter{subsubappendixc}[subappendixc]
\renewcommand{\thesubappendixc}{\Alph{appendixc}.\arabic{subappendixc}}
\renewcommand{\thesubsubappendixc}
        {\Alph{appendixc}.\arabic{subappendixc}.\arabic{subsubappendixc}}

\renewcommand{\appendix}[1] {\vspace{12pt}
        \refstepcounter{appendixc}
        \setcounter{figure}{0}
        \setcounter{table}{0}
        \setcounter{lemma}{0}
        \setcounter{theorem}{0}
        \setcounter{corollary}{0}
        \setcounter{definition}{0}
        \setcounter{equation}{0}
        \renewcommand{\thefigure}{\Alph{appendixc}.\arabic{figure}}
        \renewcommand{\thetable}{\Alph{appendixc}.\arabic{table}}
        \renewcommand{\theappendixc}{\Alph{appendixc}}
        \renewcommand{\thelemma}{\Alph{appendixc}.\arabic{lemma}}
        \renewcommand{\thetheorem}{\Alph{appendixc}.\arabic{theorem}}
        \renewcommand{\thedefinition}{\Alph{appendixc}.\arabic{definition}}
        \renewcommand{\thecorollary}{\Alph{appendixc}.\arabic{corollary}}
        \renewcommand{\theequation}{\Alph{appendixc}.\arabic{equation}}
        \noindent{\tenbf Appendix \theappendixc #1}\par\vspace{5pt}}
\newcommand{\subappendix}[1] {\vspace{12pt}
        \refstepcounter{subappendixc}
        \noindent{\bf Appendix \thesubappendixc. {\kern1pt \bfit #1}}
        \par\vspace{5pt}}
\newcommand{\subsubappendix}[1] {\vspace{12pt}
        \refstepcounter{subsubappendixc}
        \noindent{\rm Appendix \thesubsubappendixc. {\kern1pt \tenit #1}}
        \par\vspace{5pt}}

\newcommand{\textlineskip}{\baselineskip=13pt}
\newcommand{\smalllineskip}{\baselineskip=10pt}

\newcommand{\copyrightheading}[1]
        {\vspace*{-2.5cm}\smalllineskip{\flushleft
        {\footnotesize International Journal of Modern Physics D, #1}\\
        {\footnotesize \copyright\kern2pt World Scientific Publishing
         Company}\\
         }}

\newcommand{\publisher}[2]{{\begin{center}\footnotesize\smalllineskip 
        Received #1\\
        Revised #2
        \end{center}
        }}

\def\abstracts#1#2#3{{
        \centering{\begin{minipage}{4.5in}\footnotesize\baselineskip=10pt
        \parindent=0pt #1\par 
        \parindent=15pt #2\par
        \parindent=15pt #3
        \end{minipage}}\par}} 

\def\keywords#1{{
        \centering{\begin{minipage}{4.5in}\footnotesize\baselineskip=10pt
        {\footnotesize\it Keywords}\/: #1
         \end{minipage}}\par}}

\renewenvironment{thebibliography}[1]
        {\frenchspacing
         \ninerm\baselineskip=11pt
         \begin{list}{\arabic{enumi}.}
        {\usecounter{enumi}\setlength{\parsep}{0pt}     
         \setlength{\leftmargin 12.7pt}{\rightmargin 0pt}
         \setlength{\itemsep}{0pt} \settowidth
        {\labelwidth}{#1.}\sloppy}}{\end{list}}

\newcounter{itemlistc}
\newcounter{romanlistc}
\newcounter{alphlistc}
\newcounter{arabiclistc}

\newcommand{\fcaption}[1]{
        \refstepcounter{figure}
        \setbox\@tempboxa = \hbox{\footnotesize Fig.~\thefigure. #1}
        \ifdim \wd\@tempboxa > 5in
           {\begin{center}
        \parbox{5in}{\footnotesize\smalllineskip Fig.~\thefigure. #1}
            \end{center}}
        \else
             {\begin{center}
             {\footnotesize Fig.~\thefigure. #1}
              \end{center}}
        \fi}

\newcommand{\tcaption}[1]{
        \refstepcounter{table}
        \setbox\@tempboxa = \hbox{\footnotesize Table~\thetable. #1}
        \ifdim \wd\@tempboxa > 5in
           {\begin{center}
        \parbox{5in}{\footnotesize\smalllineskip Table~\thetable. #1}
            \end{center}}
        \else
             {\begin{center}
             {\footnotesize Table~\thetable. #1}
              \end{center}}
        \fi}

\def\@citex[#1]#2{\if@filesw\immediate\write\@auxout
        {\string\citation{#2}}\fi
\def\@citea{}\@cite{\@for\@citeb:=#2\do
        {\@citea\def\@citea{,}\@ifundefined
        {b@\@citeb}{{\bf ?}\@warning
        {Citation `\@citeb' on page \thepage \space undefined}}
        {\csname b@\@citeb\endcsname}}}{#1}}

\newif\if@cghi
\def\cite{\@cghitrue\@ifnextchar [{\@tempswatrue
        \@citex}{\@tempswafalse\@citex[]}}
\def\citelow{\@cghifalse\@ifnextchar [{\@tempswatrue
        \@citex}{\@tempswafalse\@citex[]}}
\def\@cite#1#2{{$\null^{#1}$\if@tempswa\typeout
        {IJCGA warning: optional citation argument 
        ignored: `#2'} \fi}}

\def\pmb#1{\setbox0=\hbox{#1}
        \kern-.025em\copy0\kern-\wd0
        \kern.05em\copy0\kern-\wd0
        \kern-.025em\raise.0433em\box0}

\def\fnt#1#2{\footnotetext{\kern-.3em
        {$^{\mbox{\scriptsize #1}}$}{#2}}}

\def\fpage#1{\begingroup
\voffset=.3in
\thispagestyle{empty}\begin{table}[b]\centerline{\footnotesize #1}
        \end{table}\endgroup}

\def\runninghead#1#2{\pagestyle{myheadings}
\markboth{{\protect\footnotesize\it{\quad #1}}\hfill}
{\hfill{\protect\footnotesize\it{#2\quad}}}}
\font\tenrm=cmr10
\font\tenit=cmti10 
\font\tenbf=cmbx10
\font\bfit=cmbxti10 at 10pt
\font\ninerm=cmr9

\font\eightrm=cmr8

\def\qed{\hbox{${\vcenter{\vbox{                  
   \hrule height 0.4pt\hbox{\vrule width 0.4pt height 6pt
   \kern5pt\vrule width 0.4pt}\hrule height 0.4pt}}}$}}

\begin{document}
\setlength{\textheight}{7.7truein}    

\runninghead{Bianchi Type I Magnetofluid Cosmological Models With Variable 
Cosmological Constant Revisited  } 
{A. Pradhan and S. K. Singh}

\normalsize\textlineskip
\thispagestyle{empty}
\setcounter{page}{1}

\copyrightheading{}             {Vol.~0, No.~0 (2003) 000--000}

\vspace*{0.88truein}

\fpage{1}

\centerline{\bf BIANCHI TYPE I MAGNETOFLUID COSMOLOGICAL MODELS WITH}
\vspace*{0.035truein}
\centerline{\bf VARIABLE COSMOLOGICAL CONSTANT REVISITED}
\vspace*{0.37truein}
\centerline{\footnotesize ANIRUDH PRADHAN\footnote{(Corresponding Author)}}
\vspace*{0.015truein}  
\centerline{\footnotesize\it Department of Mathematics, Hindu Post-graduate College,}
\baselineskip=10pt
\centerline{\footnotesize\it Zamania, Ghazipur 232 331, India}
\vspace*{0.015truein} 
\centerline{\footnotesize\it acpradhan@yahoo.com, pradhan@iucaa.ernet.in}
\vspace*{10pt}
\centerline{\footnotesize SANJAY KUMAR SINGH}
\centerline{\footnotesize\it Department of Computer Science, Post-graduate College, }
\vspace*{0.015truein}
\centerline{\footnotesize\it Ghazipur 233 001, India}
\vspace*{0.015truein}
\centerline{\footnotesize\it sanjaysingh10@rediffmail.com}
\baselineskip=10pt
\vspace*{0.225truein}
\publisher{(received date)}{(revised date)}
\vspace*{0.21truein}
\abstracts{The behaviour of magnetic field in anisotropic Bianchi type I cosmological model
for bulk viscous distribution is investigated. The distribution consists of an 
electrically neutral viscous fluid with an infinite electrical conductivity. It is 
assumed that the component $\sigma^{1}_{1}$ of shear tensor $\sigma^{j}_{i}$ is 
proportional to expansion ($\theta$) and the coefficient of bulk viscosity is assumed 
to be a power function of mass density. Some physical and geometrical aspects  of the 
models are also discussed in presence and also in absence of the magnetic field.}{}{}


\vspace*{1pt}\textlineskip      
\vspace*{-0.5pt}
\section{Introduction}
A resurgence of interest in anisotropic, general-relativistic cosmological
models of the universe \cite{ref1,ref2} has been stimulated by the discovery
of the cosmic microwave radiation, the study of its isotropy, the problem of 
the abundance of primordial helium, and the possibility of large-scale primordial
magnetic fields. The choice of anisotropic cosmological models in Einstein system
of field equations leads to the cosmological models more general than 
Robertson-Walker model.\cite{ref3} The occurrence of magnetic fields on galactic 
scale is well-established fact today, and their importance for a variety of 
astrophysical phenomena is generally acknowledged as pointed out by Zeldovich 
et al.\cite{ref4} Also Harrison \cite{ref5} has suggested that magnetic field 
could have a cosmological origin. As a natural consequences we should include 
magnetic fields in the energy-momentum tensor of the early universe.
The presence of primordial magnetic fields in the early stages of the evolution 
of the universe has been discussed by several authors.\cite{ref6} $^-$\cite{ref15} 
Strong magnetic fields can be created due to adiabatic compression in clusters of 
galaxies. Large-scale magnetic fields give rise to anisotropies in the universe. 
The  anisotropic pressure created by the magnetic fields dominates the evolution 
of the shear anisotropy and it decays slower than if the pressure was 
isotropic.\cite{ref16,ref17} Such fields can be generated at the end of an inflationary 
epoch.\cite{ref18}$^-$\cite{ref22} Anisotropic magnetic field models have 
significant contribution in the evolution of galaxies and stellar objects. 
Bali and Ali\cite{ref23} obtained a magnetized cylindrically symmetric universe 
with an electrically neutral perfect fluid as the source of matter. Several authors
\cite{ref24}$^-$\cite{ref28} have investigated Bianchi type I cosmological 
models with a magnetic field in different context.\\
\newline
\par
Most cosmological models assume that the matter in the universe can be described 
by `dust'(a pressure-less distribution) or at best a perfect fluid. Nevertheless,
there is good reason to believe that - at least at the early stages of the universe 
- viscous effects do play a role.\cite{ref27}$^-$\cite{ref29} For example, the 
existence of the bulk viscosity is equivalent to slow process of restoring 
equilibrium states.\cite{ref30} The observed physical phenomena such as the 
large entropy per baryon and remarkable degree of isotropy of the cosmic 
microwave background radiation suggest analysis of dissipative effects in cosmology.
Bulk viscosity is associated with the GUT phase transition and string creation. 
Thus, we should consider the presence of a material distribution other than a 
perfect fluid to have realistic cosmological models (see Gr\o n\cite{ref31} for 
a review on cosmological models with bulk viscosity). The effect of bulk viscosity 
on the cosmological evolution has been investigated by a number of authors in the 
framework of general theory of relativity.\cite{ref32}$^-$\cite{ref46}\\  
\newline
\par
In recent years, models with relic cosmological constant $\Lambda$ have drawn
considerable attention among researchers for various aspects such as the age
problem, classical tests, observational constraints on $\Lambda$, structure
formation and gravitational lenses have been discussed in the literature.
It is remarkable here that in the absence of any interaction with matter or 
radiation this would force the cosmological constant to be constant, but, in 
the presence of the interaction with matter or radiation, a solution of Einstein's
field equation and assumed equation of covariant conservation of energy with 
a time varying $\Lambda$ can be found. For these solutions, conservation of 
energy requires that any decrease in the energy density of the vacuum component 
be compensated for by a corresponding increase in the energy density of matter 
or radiation. Some of the recent discussions on the cosmological constant
``problem'' and consequence on cosmology with a time-varying cosmological
constant are contained in Bertolami,\cite{ref47} Ratra and Peebles,\cite{ref48}
Dolgov,\cite{ref49}$^-$\cite{ref51} Sahni and Starobinsky,\cite{ref52} 
Padmanabhan,\cite{ref53} Vishwakarma.\cite{ref54} Recent observations by 
Perlmutter {\it et al.} \cite{ref55} and Riess {\it et al.} \cite{ref56} 
strongly favour a significant and positive $\Lambda$. Their finding arise from 
the study of more than $50$ type Ia supernovae with redshifts in the range 
$0.10 \leq z \leq 0.83$ and suggest Friedmann models with negative pressure 
matter such as a cosmological constant, domain walls or cosmic strings 
(Vilenkin,\cite{ref57} Garnavich{\it et al.}\cite{ref58}) Recently, 
Carmeli and Kuzmenko\cite{ref59} have shown that the cosmological relativity 
theory (Behar and Carmeli\cite{ref60}) predicts the value 
$\Lambda = 1.934\times 10^{-35} s^{-2}$ for the cosmological 
constant. This value of $\Lambda$ is in excellent agreement with the measurements 
recently obtained by the High-Z Supernova Team and Supernova Cosmological Project 
(Garnavich {\it et al.};\cite{ref58} Perlmutter {\it et al.};\cite{ref55} 
Riess {\it et al.};\cite{ref56} Schmidt {\it et al.}\cite{ref61}) The main 
conclusion of these works is that the expansion of the universe is accelerating.   
\newline
\par
Recently, Bali and Jain\cite{ref62} have obtained a generalized expanding and 
shearing anisotropic Bianchi type I magnetofluid cosmological model for perfect 
fluid distribution in general relativity. Motivated by the situations discussed 
above in regard with bulk viscous cosmologies, we extend their work by including an
electrically neutral bulk viscous fluid as the source of matter in the energy-
momentum tensor. The paper is organized as follows. In Sec. 2, we review the
solutions and the main results of Bali and Jain.\cite{ref62} Here we have shown
that all the solutions obtained by Bali and Jain\cite{ref62} and Bali\cite{ref71}
can be derived from our solutions. In Sec. 3, we obtain the solutions for a universe 
filled with a bulk viscosity in presence of magnetic field. Sec. 4 includes the bulk 
viscous cosmological solutions in the absence of magnetic field. In Sec. 5, we discuss 
our main results and summarize our conclusions.\\ 
\section{A Magnetic Fluid Universe-Revisited}
In this section, we review the solutions obtained by Bali and Jain.\cite{ref62} We
consider an anisotropic homogeneous Bianchi type I metric in the form given by
Marder\cite{ref63}
\begin{equation} 
\label{eq1}  
ds^{2} = A^{2}(dx^{2} - dt^{2}) + B^{2}dy^{2} + C^{2}dz^{2},
\end{equation} 
where the metric potentials are functions of $t$ only. The energy momentum tensor
is taken into the form
\begin{equation} 
\label{eq2}
T^{j}_{i} = (\rho + \bar{p})v_{i}v^{j} + \bar{p} g^{j}_{i} + E^{j}_{i},
\end{equation} 
where $E^{j}_{i}$ is the electro-magnetic field given by Lichnerowicz\cite{ref64}
as 
\begin{equation} 
\label{eq3}
E^{j}_{i} = \bar{\mu}\left[|h|^{2}\left(v_{i}v^{j} + \frac{1}{2}g^{j}_{i}\right)
- h_{i}h^{j}\right],
\end{equation} 
and
\begin{equation} 
\label{eq4}
\bar{p} = p - \xi v^{i}_{;i}.
\end{equation} 
Here $\rho$, $p$ $\bar{p}$ and $\xi$ are the energy density, isotropic pressure, effective 
pressure, bulk viscous coefficient respectively and $v^{i}$ is the 
flow vector satisfying the relation
\begin{equation} 
\label{eq5}
g_{ij} v^{i}v^{j} = - 1.
\end{equation} 
$\bar\mu$ is the magnetic permeability and $h_{i}$ the magnetic flux vector
defined by
\begin{equation} 
\label{eq6}
h_{i} = \frac{1}{\bar{\mu}}~~ ^*F_{ji}v^{j},
\end{equation} 
where $^*F_{ij}$ is the dual electro-magnetic field tensor defined by Synge
\cite{ref65} to be
\begin{equation} 
\label{eq7}
^*F_{ij} = \frac{\sqrt-g}{2}\epsilon_{ijkl} F^{kl}.
\end{equation} 
$F_{ij}$ is the electro-magnetic field tensor and $\epsilon_{ijkl}$ is the
Levi-Civita tensor density. Here, the comoving coordinates are taken to be 
$v^{1}$ = $0$ = $v^{2}$ = $v^{3}$ and $v^{4}$ = $\frac{1}{A}$. We take the 
incident magnetic field to be in the direction of $x$-axis so that $h_{1} 
\ne 0$, $h_{2}$ =  $0$ = $h_{3}$ = $h_{4}$. This leads to $F_{12} = 0 = F_{13}$ 
by virtue of (\ref{eq6}). Also due to assumption of infinite conductivity of the 
fluid, we get $F_{14} = 0 = F_{24} = F_{34}$. The only non-vanishing component 
of $F_{ij}$ is $F_{23}$. The first set of Maxwell's equation
\begin{equation} 
\label{eq8}
F_{ij;k} + F_{jk;i} + F_{ki;j} = 0,
\end{equation}  
leads to
\begin{equation} 
\label{eq9} 
F_{23} = I ~ ~ \mbox{(const.)},
\end{equation}  
where the semicolon represents a covariant differentiation. Hence
\begin{equation} 
\label{eq10} 
h_{1} = \frac{AI}{\bar{\mu}BC}.
\end{equation}  
The Einstein's field equations with time-dependent cosmological constant  
\begin{equation} 
\label{eq11} 
R^{j}_{i} - \frac{1}{2} R g^{j}_{i} + \Lambda g^{j}_{i} = - 8\pi T^{j}_{i}, ~ ~ 
\mbox{(c = 1, G = 1 in gravitational unit)}
\end{equation}  
for the line element (1) has been set up as
\begin{equation} 
\label{eq12} 
8\pi A^{2}\left(\bar{p} - \frac{I^{2}}{2\bar{\mu}B^{2}C^{2}}\right) = -\frac{B_{44}}{B}
- \frac{C_{44}}{C} - \frac{B_{4}C_{4}}{BC} + \frac{A_{4} B_{4}}{AB} + 
\frac{A_{4}C_{4}}{AC}- \Lambda A^{2},
\end{equation}  
\begin{equation} 
\label{eq13} 
8\pi A^{2} \left(\bar{p} + \frac{I^{2}}{2\bar{\mu}B^{2}C^{2}}\right) = -\frac{A_{44}}{A}
- \frac{C_{44}}{C} + \frac{A^{2}_{4}}{A^{2}} - \Lambda A^{2},
\end{equation}  
\begin{equation} 
\label{eq14} 
8\pi A^{2} \left(\bar{p} + \frac{I^{2}}{2\bar{\mu}B^{2}C^{2}}\right) = -\frac{A_{44}}{A}
- \frac{B_{44}}{B} + \frac{A^{2}_{4}}{A^{2}} - \Lambda A^{2},
\end{equation}  
\begin{equation} 
\label{eq15} 
8\pi A^{2} \left(\rho +  \frac{I^{2}}{2\bar{\mu}B^{2}C^{2}}\right) = \frac{A_{4}B_{4}}{AB}
+ \frac{A_{4}C_{4}}{AC} + \frac{B_{4}C_{4}}{BC} + \Lambda A^{2}.
\end{equation}  
The suffix $4$ by the symbols $A$, $B$ and $C$ denote differentiation with
respect to $t$. Equations (\ref{eq12})-(\ref{eq15}) are four equations in 
seven unknowns, $A$, $B$, $C$, $\rho$, $p$, $\xi$ and $\Lambda$.\\
Referring to Thorn,\cite{ref66} current observations of the velocity-redshift relation 
for extragalactic sources suggest that Hubble expansion of the universe is isotropic 
today to within $\sim 30$ per cent.\cite{ref67,ref68} Put more precisely, redshift
studies place the limit
\[
\frac{\sigma}{H}\leq 0.3
\]
on the ratio of shear, $\sigma$, to Hubble constant, $H$, in the neighbourhood of
of our Galaxy today. \\
Following Bali and Jain,\cite{ref62} we assume that the expansion ($\theta$) 
in the model is proportional to the eigen value $\sigma^{1}_{1}$ of the shear tensor 
$\sigma^{j}_{i}$. This condition leads to
\begin{equation} 
\label{eq16} 
A = (BC)^{n},
\end{equation}  
where $n$ is the proportionality constant and
\[
\sigma^{1}_{1} = \frac{1}{3A}\left(\frac{2A_{4}}{A} - \frac{B_{4}}{B} - \frac{C_{4}}{C}\right)
\]
\[
\theta =  \frac{1}{A}\left(\frac{A_{4}}{A} + \frac{B_{4}}{B} + \frac{C_{4}}{C}\right)
\]
for the metric (\ref{eq1}). Solving these fields equations, we obtain
\begin{equation} 
\label{eq17} 
A^{2} = \frac{\sin^{2} (at)}{K_{1}},  
\end{equation}
\begin{equation} 
\label{eq18} 
B^{2} = \frac{N} {K_{1}^{\frac{1}{2n}(\frac{L}{\sqrt{M}} + 1)}}\sin^{\frac{1}{n}} (at) 
 ~ \left(\tan \frac{at}{2}\right)^{\frac {L}{n\sqrt{M}}},
\end{equation}
\begin{equation} 
\label{eq19} 
C^{2} = \frac{1}{N K_{1}^{\frac{1}{2n} (1 - \frac{L}
{\sqrt{M}})}} \sin^{\frac{1}{n}} (at) \left(\tan \frac{at}{2}\right)^
{\frac{- L}{n\sqrt{M}}},
\end{equation}
where 
$M$, $N$ and $L$ are constants of integration and
\[
a^{2} = \frac{16 \pi I^{2}}{\bar{\mu}},
\]
\[
K_{1} = \frac{a^{2}}{n(2n - 1)M}.
\]
Hence the metric (\ref{eq1}) reduces to the form
\[
ds^{2} =  \frac{\sin^{2} (at)}{K_{1}}\left[dx^{2} - \frac{K_{2}^{\frac{1}{n}}}
{n^{2} M a^{\frac{2(1- n)}{n}}} \sin^{\frac{2(1 - n)}{n}} (at) dt^{2}\right]
\]
\[
+ \frac{N} {K_{1}^{\frac{1}{2n}(\frac{L}{\sqrt{M}} + 1)}}\sin^{\frac{1}{n}} (at) 
\left[\tan \frac{at}{2}\right]^{\frac {L}{n\sqrt{M}}} dy^{2} 
\]
\begin{equation} 
\label{eq20}
+ \frac{1}{N K_{1}^{\frac{1}{2n} (1 - \frac{L}
{\sqrt{M}})}}\sin^{\frac{1}{n}} (at) \left[\tan \frac{at}{2}\right]^
{\frac{- L}{n\sqrt{M}}}dz^{2},
\end{equation}
where
\[
K_{2} = n(2n - 1) M.
\]
Using the transformations \\
$
x = X, \\
\sqrt{N} y = Y, \\
\frac{z}{\sqrt{N}} = Z, \\
t = T, \\
$
the metric (\ref{eq20}) reduces to the form
\[
ds^{2} =  \frac{\sin^{2} (aT)}{K_{1}}\left[dX^{2} - \frac{K_{2}^{\frac{1}{n}}}
{n^{2} M a^{\frac{2(1- n)}{n}}} \sin^{\frac{2(1 - n)}{n}} (a T) dT^{2}\right]
\]
\[
+  \frac{N} {K_{1}^{\frac{1}{2n}(1 + \beta)}} \sin^{\frac{1}{n}}( a T) 
 \left[\tan \frac{a T}{2}\right]^{\frac{\beta}{n}} dY^{2} 
\]
\begin{equation} 
\label{eq21}
+ \frac{1}{N K_{1}^{\frac{1}{2n} (1 - \beta)}} \sin^{\frac{1}{n}}( a T)
\left[\tan \frac{a T}{2}\right]^{\frac{- \beta}{n}}dZ^{2},
\end{equation}
where
\[
\beta = \frac{L}{\sqrt{M}}.
\]
The effective pressure and energy density for the model (\ref{eq21}) are given by
\begin{equation} 
\label{eq22}
8\pi \bar{p} = \frac{K_{1}^{\frac{n + 1}{n}} \left[K_{3} - L^{2} + (2 K_{2} - M)\sin^{2} (aT)\right]}
{4 \sin^{\frac{2(n + 1)}{n}} (aT)} - \frac{a^{2} K_{1}^{\frac{1}{n}}}{4\sin^{\frac{2}{n}} (a T)}
- \Lambda,
\end{equation}
\begin{equation} 
\label{eq23}
8\pi \rho  = \frac{K_{1}^{\frac{n + 1}{n}} \left[K_{3}\cos^{2} (aT) - L^{2}\right]}
{4 \sin^{\frac{2(n + 1)}{n}} (aT)} - \frac{a^{2} K_{1}^{\frac{1}{n}}}{4\sin^{\frac{2}{n}} (a T)}
+ \Lambda,
\end{equation}
where 
$
K_{3} = (4n + 1)M 
$.\\
In the absence of magnetic field, the metric (\ref{eq20}) reduces to
\[
ds^{2} = K_{2} T^{2}\left[dX^{2} - \frac{K_{2}^{\frac{1}{n}}}{n^{2} M} 
T^{\frac{2(1 - n)}{n}}dT^{2}\right] 
\]
\begin{equation} 
\label{eq24}
+ \frac{N K_{2}^{\frac{(1 + \beta)}{2n}}} {2^{\frac{\beta}{n}}}
T^{\frac{1 + \beta}{n}} dY^{2} + \frac{K_{2}^{\frac{(1 - \beta)}{2n}}}
{N 2^{\frac{-\beta}{n}}} T^{\frac{1 - \beta}{n}} dZ^{2}. 
\end{equation}
The effective pressure and energy density for the model (\ref{eq24}) are given by
\begin{equation} 
\label{eq25}
8\pi \bar{p} = \frac{[K_{3} - L^{2} + (2K_{2} - M) T^{2}]}{4(K_{2} T^{2})^
{\frac{n + 1}{n}}}  - \frac{1}{4 (K_{2} T^{2})^{\frac{1}{n}}}  - \Lambda,
\end{equation}
\begin{equation} 
\label{eq26}
8\pi \rho  = \frac{(K_{3} - L^{2})}{4(K_{2}T^{2})^{\frac{n + 1}{n}}} 
- \frac{1}{4(K_{2} T^{2})^{\frac{1}{n}}} + \Lambda,
\end{equation}
If we consider $\Lambda$ as constant and coefficient of bulk viscosity $\xi$  as zero, 
we obtain the solutions obtained by Bali and Jain.\cite{ref62} If we set $\Lambda = constant$, 
$\xi = 0$ and $n = 1$, we get the solutions obtained by Bali.\cite{ref71} \\
\section{Bulk Viscous Solutions in the Presence of a Magnetic Field}
The equations with bulk viscosity can be obtained from the general relativistic
field equations as follows.\cite{ref37} To obtain these equations, we replace
$p$ by the effective pressure $\bar p$ given by (\ref{eq4}).
The expansion scalar  $\theta$ is given by
\begin{equation}
\label{eq27}
\theta = \sqrt{M}~ (n + 1) K_{1}^{\frac{n + 1}{2n}}~ \frac{\cos(aT)}{\sin^{\frac{n + 1}
{n}} (aT)}
\end{equation} 
Eq. (\ref{eq22}) can be rewritten as
\begin{equation} 
\label{eq28}
8\pi (p - \xi \theta) = \frac{K_{1}^{\frac{n + 1}{n}} \left[K_{3} - L^{2} + (2 K_{2} - M)
\sin^{2} (aT)\right]}{4 \sin^{\frac{2(n + 1)}{n}} (aT)} - \frac{a^{2} K_{1}^{\frac{1}{n}}}
{4\sin^{\frac{2}{n}} (a T)} - \Lambda,
\end{equation}
For the specification of $\xi$, now we assume that the fluid obeys an equation
of state of the form
\begin{equation} 
\label{eq29} 
p = \gamma \rho,
\end{equation} 
where $\gamma(0\leq\gamma \leq 1)$ is a constant.\\
Thus, given $\xi(t)$ we can solve the cosmological parameters. In most of the 
investigations involving bulk viscosity is assumed to be a simple power function of 
the energy density.\cite{ref34}$^-$\cite{ref36} 
\begin{equation}
\label{eq30}
\xi(t) = \xi_{0} \rho^{m},
\end{equation}
where $\xi_{0}$ and $m$ are constants. If $m = 1$, Eq. (\ref{eq30}) may correspond
to a radiative fluid.\cite{ref37} However, more realistic models\cite{ref38} are 
based on $m$ lying in the regime $0 \leq m \leq \frac{1}{2}$. \\
\subsection {Model I: ~ ~ ~ $(\xi = \xi_{0})$}
When $m = 0$, Eq. (\ref{eq30}) reduces to $\xi$ = $\xi_{0}$ and hence 
Eqs. (\ref{eq28}), with the use of Eq. (\ref{eq22}), (\ref{eq27}) and (\ref{eq29}) 
lead to
\[
8\pi (1 + \gamma)\rho = \frac{8\pi \xi_{0} (n + 1) \sqrt{M} K_{1}^{\frac{n + 1}{2n}}
\cos (aT)}{\sin^{\frac{n + 1}{n}}(aT)} + 
\]
\begin{equation} 
\label{eq31}
\frac{K_{1}^{\frac{n + 1}{n}} \left[2(K_{3} - L^{2}) + 2K_{4}M \sin^{2} 
(aT)\right]}{4 \sin^{\frac{2(n + 1)}{n}} (aT)} 
- \frac{a^{2} K_{1}^{\frac{1}{n}}}{2\sin^{\frac{2}{n}} (a T)},
\end{equation} 
Eliminating $\rho(t)$ between Eqs. (\ref{eq22}) and (\ref{eq31}), we obtain
\[
(1 + \gamma)\Lambda = \frac{8\pi \xi_{0} (n + 1) \sqrt{M} K_{1}^{\frac{n + 1}{2n}}
\cos (aT)}{\sin^{\frac{n + 1}{n}} (aT)} + 
\frac{K_{1}^{\frac{n + 1}{n}} \left[K_{3} - L^{2} + (2K_{3} - M) \sin^{2} 
(aT)\right]}{4 \sin^{\frac{2(n + 1)}{n}} (aT)} - 
\]
\begin{equation} 
\label{eq32}
\frac{K_{1}^{\frac{n + 1}{n}} \left[K_{3}\cos^{2}(aT) - L^{2}\right]\gamma}
{4 \sin^{\frac{2(n + 1)}{n}} (aT)} + \frac{a^{2} K_{1}^{\frac{1}{n}} \gamma}
{2\sin^{\frac{2}{n}} (a T)},
\end{equation}
where $K_{4} = 2n^{2} - 3n -1$. \\
\subsection {Model II: ~ ~ ~ $(\xi = \xi_{0}\rho)$}
When $m = 1$, Eq. (\ref{eq30}) reduces to $\xi = \xi_{0}\rho$ and hence 
Eqs. (\ref{eq28}), with the use of Eq. (\ref{eq23}), (\ref{eq27}) and (\ref{eq29}) 
lead to
\[
8 \pi \rho = \frac{1}{\left[1 + \gamma - \frac{\xi_{0} (n + 1) \sqrt{M} K_{1}^
{\frac{n + 1}{2n}}\cos (aT)}{\sin^{\frac{(n + 1)}{n}} (aT)}\right]} \times
\]
\begin{equation} 
\label{eq33}
\left[\frac{K_{1}^{\frac{n + 1}{n}} \left[2(K_{3} - L^{2}) + 2K_{4}
M\sin^{2} (aT)\right]}{4 \sin^{\frac{2(n + 1)}{n}} (aT)} -  \frac{a^{2} K_{1}^
{\frac{1}{n}}}{2\sin^{\frac{2}{n}} (a T)}\right],
\end{equation}
Eliminating $\rho(t)$ between Eqs. (\ref{eq23}) and (\ref{eq33}), we obtain
\[
\Lambda  = \frac{1}{\left[1 + \gamma - \frac{\xi_{0} (n + 1) \sqrt{M} K_{1}^
{\frac{n + 1}{2n}}\cos (aT)}{\sin^{\frac{(n + 1)}{n}} (aT)}\right]} \times
\]
\[
\left[\frac{K_{1}^{\frac{n + 1}{n}} \left[2(K_{3} - L^{2}) + 2K_{4}
M\sin^{2} (aT)\right]}{4 \sin^{\frac{2(n + 1)}{n}} (aT)} -  \frac{a^{2} K_{1}^
{\frac{1}{n}}}{2\sin^{\frac{2}{n}} (a T)}\right]
\]
\begin{equation} 
\label{eq34}
- \left[\frac{K_{1}^{\frac{n + 1}{n}}\left(K_{3} \cos^{2} (aT)  - L^{2}\right)}
 {4 \sin^{\frac{2(n + 1)}{n}} (aT)} -  \frac{a^{2} K_{1}^
{\frac{1}{n}}}{4\sin^{\frac{2}{n}} (a T)}\right].
\end{equation}
{\bf Some Physical and Geometrical Features  of the Models}\\
We shall now give the expressions for kinematical quantities and the components
of conformal curvature tensor. The scalar of expansion $\theta$ calculated for 
the flow vector $v^{i}$ is already given by (\ref{eq27}).\\
The rotation $\omega$ is identically zero and the shear in the model, is given by
\begin{equation} 
\label{eq35} 
\sigma^{2} = \frac{K_{1}^{\frac{n + 1}{n}}[(2n - 1)^{2} M \cos^{2}(aT) + 3 L^{2}]}
{12 \sin^{\frac{2(n + 1)}{n}}(aT)}
\end{equation}
The non-vanishing components of conformal curvature tensor are obtained as
\begin{equation} 
\label{eq36} 
C^{12}_{12} = - \frac{K_{1}^{\frac{n + 1}{n}}[L^{2} - M + 2n M + 6 n L \sqrt{M}
\cos(aT) + K_{4}M \sin^{2} (aT)]}
{12 \sin^{\frac{2(n + 1)}{n}}(aT)},
\end{equation}
\begin{equation} 
\label{eq37} 
C^{13}_{13} = - \frac{K_{1}^{\frac{n + 1}{n}}[L^{2} - M + 2n M - 6 n L \sqrt{M}
\cos(aT) + K_{4}M \sin^{2} (aT)]}
{12 \sin^{\frac{2(n + 1)}{n}}(aT)},
\end{equation}
\begin{equation} 
\label{eq38} 
C^{14}_{14} =  \frac{K_{1}^{\frac{n + 1}{n}}[L^{2} - M + 2n M  + K_{4}M
\sin^{2} (aT)]}{12 \sin^{\frac{2(n + 1)}{n}}(aT)}.
\end{equation}
The model represents an expanding, shearing and non-rotating universe in general.
The Eq. (\ref{eq21}) requires that $n(2n - 1) \neq 0$. If we choose $n > 0$, then
$\rho$ and $p$ becomes infinite at $T = 0$ and $T = \frac{\pi}{a}$. Eq. (\ref{eq27})
shows that $\theta = \infty$ at $T = 0$, $\theta = 0$ at $T = \frac{\pi}{2a}$ and
$\theta = - \infty$ at $T = \frac{\pi}{a}$. Thus the model expands with a big bang
at $T = 0$, the expansion stops at $T = \frac{\pi}{2a}$ and collapses at 
$T = \frac{\pi}{a}$ much like the closed FRW model. The magnetic field is responsible 
for this.\\
It is observed that the metric exhibits singularities at $T = 0$ and 
$T = \frac{\pi}{a}$. At $T = 0$ and $T = \frac{\pi}{a}$, the model (\ref{eq21}) has 
singularity of Point type, Barrel type and Cigar type if $\beta < 1$, $\beta = 1$ and 
$\beta > 1$ respectively as given by MacCallum \cite{ref69} and Islam.\cite{ref70} 
At the singularity $T = 0$, $g_{22} \rightarrow \infty $, $g_{33} \rightarrow \infty $
according as $\frac{1 + \beta}{n} < 0$ and $\frac{1 - \beta}{n} < 0$. Also at 
$T = \frac{\pi}{a}$, $g_{22} \rightarrow \infty $, $g_{33} \rightarrow \infty $
according as $\frac{1 - \beta}{n} < 0$ and $\frac{1 + \beta}{n} < 0$. Thus  inflationary 
scenario exists near the initial singularities $T= 0$ and $T = \frac{\pi}{a}$ in the model. \\
The cosmic time $t$ is given by 
\begin{equation} 
\label{eq39} 
t = k_{1} \int \sin^{\frac{1}{n}} (aT) dT,
\end{equation}
where $k_{1}$ is constant. Near $T =0$, $\sin^{\frac{1}{n}} (aT) \simeq a^{\frac{1}{n}}
T^{\frac{1}{n}}$ i.e., $t \simeq k_{2} T^{\frac{1}{n}}$, where $k_{2}$ is constant. Thus
$T \rightarrow 0$ implies $t \rightarrow 0$. When $T \rightarrow \frac{\pi}{a}$ then
$t \rightarrow constant$. Hence the model has a finite time span in cosmic time scale.
The space time is Petrov type D when $L = 0$ and non-degenerate Petrov type I otherwise.\\
The expressions for $\frac{E^{4}_{4}}{\rho}$ = $\frac{magnetic  ~ energy}{material ~ energy}$,
$\frac{\sigma}{\theta}$ are given as follows:
\begin{equation} 
\label{eq40} 
\frac{E^{4}_{4}}{\rho} = \frac{- a^{2} \sin^{2} (aT)}{K_{1}[K_{3}\cos^{2} (aT) - L^{2}]
-[a^{2} - \frac{\Lambda \sin^{\frac{2}{n}} (aT)}{K^{\frac{1}{n}}_{1}}] \sin^{\frac{2}{n}} 
(aT)} 
\end{equation}
\begin{equation} 
\label{eq41} 
\frac{\sigma}{\theta} = \frac{1}{2\cos(aT)}\left[\frac{(2n -1)^{2}\cos^{2}(aT) +
\frac{3L^{2}}{M}}{3(n + 1)^{2}}\right]^{\frac{1}{2}}. 
\end{equation}
It is observed that when $T \rightarrow 0$, then $\frac{E_{44}}{\rho} \rightarrow 0$,
which shows that the material energy is more dominant than magnetic near the initial
singularity.\\
\section{Bulk Viscous Solutions in the Absence of Magnetic Field}
In the absence of the magnetic field, the pressure and density for model (\ref{eq24})
are given by Eqs. (\ref{eq25}) and (\ref{eq26}) respectively. Eq. (\ref{eq25}) can be 
rewritten as 
\begin{equation} 
\label{eq42} 
8\pi (p - \xi \theta) = \frac{[K_{3} - L^{2} + (2K_{2} - M) T^{2}]}{4(K_{2} T^{2})^
{\frac{n + 1}{n}}}  - \frac{1}{4 (K_{2} T^{2})^{\frac{1}{n}}}  - \Lambda,
\end{equation}
where the expansion scalar $(\theta)$  is given by
\begin{equation} 
\label{eq43} 
\theta = \frac{\sqrt{M}~ (n + 1)}{(K_{2} T^{2})^{\frac {n + 1}{2n}}}.
\end{equation}
Using Eqs. (\ref{eq29}), (\ref{eq30}) and (\ref{eq43}) in (\ref{eq42}), we obtain
\[
8 \pi \left[\gamma \rho - \xi_{0} \rho^{n} \frac{\sqrt{M}~ (n + 1)}{(K_{2} T^{2})^
{\frac {n + 1}{2n}}}\right] = \frac{[K_{3} - L^{2} + (2K_{2} - M) T^{2}]}{4(K_{2} T^{2})^
{\frac{n + 1}{n}}}  
\]
\begin{equation} 
\label{eq44} 
- \frac{1}{4 (K_{2} T^{2})^{\frac{1}{n}}}  - \Lambda, 
\end{equation}
\subsection{Model I: $ (\xi = \xi_{0})$}
When $m$ = $0$, Eq. (\ref{eq30}) reduces to $\xi$ = $\xi_{0}$ and hence
Eq. (\ref{eq44}) with the use of Eq. (\ref{eq26}) leads to
\begin{equation} 
\label{eq45} 
8\pi(1 + \gamma)\rho = \frac{8\pi \xi_{0}(n + 1)\sqrt{M}}{(K_{2}T^{2})^{\frac{n + 1}{2n}}}
+ \frac{2(K_{3} - L^{2}) + (2K_{2} - M)T^{2}}{4(K_{2}T^{2})^{\frac{n +1}{n}}} -
\frac{1}{2(K_{2}T^{2})^{\frac{1}{n}}}
\end{equation}
Eliminating $\rho(t)$ between Eqs. (\ref{eq26}) and (\ref{eq45}), we obtain
\[
(1 + \gamma)\Lambda = \frac{8\pi \xi_{0}(n + 1)\sqrt{M}}{(K_{2}T^{2})^{\frac{n +1}{2n}}}
+ \frac{K_{3} - L^{2} + (2K_{2} - M)T^{2}}{4(K_{2}T^{2})^{\frac{n +1}{n}}} 
\]
\begin{equation} 
\label{eq46} 
- \frac{(K_{3} - L^{2})\gamma}{4(K_{2}T^{2})^{\frac{n +1}{n}}} 
+ \frac{\gamma}{4(K_{2}T^{2})^{\frac{1}{n}}}
\end{equation}
\subsection{Model II: $ (\xi = \xi_{0}\rho)$ }
When $m$ = $1$, Eq. (\ref{eq30}) reduces to $\xi$ = $\xi_{0}\rho$ and hence 
Eq. (\ref{eq44}) with the help of Eq. (\ref{eq26}) lead to
\[
8\pi \rho = \frac{1}{\left[1 + \gamma - \frac{\xi_{0}(n + 1)\sqrt{M}}
{(K_{2}T^{2})^{\frac{n +1}{2n}}}\right]}\times
\] 
\begin{equation} 
\label{eq47} 
\left[\frac{2(K_{3} - L^{2}) + (2K_{2} - M)T^{2}}{4(K_{2}T^{2})^{\frac{n +1}{n}}} -
\frac{1}{2(K_{2}T^{2})^{\frac{1}{n}}}\right]
\end{equation}
Eliminating $\rho(t)$ between Eqs. (\ref{eq26}) and (\ref{eq47}), we obtain
\[
\Lambda  = \frac{1}{\left[1 + \gamma - \frac{\xi_{0}(n + 1)\sqrt{M}}
{(K_{2}T^{2})^{\frac{n +1}{2n}}}\right]}\times
\] 
\[
\left[\frac{2(K_{3} - L^{2}) + (2K_{2} - M)T^{2}}{4(K_{2}T^{2})^{\frac{n +1}{n}}} -
\frac{1}{2(K_{2}T^{2})^{\frac{1}{n}}}\right]
\]
\begin{equation} 
\label{eq48} 
- \frac{(K_{3} - L^{2})}{4(K_{2}T^{2})^{\frac{n +1}{n}}} 
+ \frac{1}{4(K_{2}T^{2})^{\frac{1}{n}}}
\end{equation}
From Eqs. (\ref{eq46}) and (\ref{eq48}), we observe that the cosmological constant 
in both models is a decreasing function of time and it approaches a small value as 
time progresses (i.e., the present epoch). The values of cosmological ``constant''
for both models are found to be small and positive which are supported by the 
results from recent supernovae observations (Garnavich {\it et al.};\cite{ref58}
 Perlmutter{\it et al.};\cite{ref55} Riess {\it et al.};\cite{ref56} Schmidt 
{\it et al.}\cite{ref61} )\\
{\bf Some Physical and Geometric Aspects of Models}\\
In absence of the magnetic field, the expansion $\theta$ is already given by
(\ref{eq43}). The expansion in the model stops for large values of $T$. The
shear in the model is given by
\begin{equation} 
\label{eq49} 
\sigma^{2} = \frac{(2n - 1)^{2}M + 3L^{2}}{12K^{\frac{n + 1}{n}}_{1}(aT)^
{\frac{2(n + 1)}{n}}}
\end{equation}
The components of conformal curvature tensor are given by
\begin{equation} 
\label{eq50} 
C^{11}_{11} = \frac{L^{2} - M + 2n M + 6n L \sqrt{M}}{12(K_{2}T^{2})^
{\frac{n + 1}{n}}},
\end{equation}
\begin{equation} 
\label{eq51} 
C^{22}_{22} = \frac{L^{2} - M + 2n M - 6n L \sqrt{M}}{12(K_{2}T^{2})^
{\frac{n + 1}{n}}},
\end{equation}
\begin{equation} 
\label{eq52} 
C^{14}_{14} = \frac{L^{2} - M + 2n M}{6(K_{2}T^{2})^{\frac{n + 1}{n}}},
\end{equation}
Thus, in the absence of magnetic field, the space-time is Petrov type $D$ if 
$L = 0$ and non-degenerate Petrov type I otherwise. For large values of $T$
, the space-time is coformally flat.  For $n > 0$, when $T$ is large, the shear 
dies out.\\
\section{Conclusions} 
We have obtained a new class of Bianchi type I anisotropic magnetofluid  cosmological
models with a bulk viscous fluid as the source of matter. Generally, the models are
expanding, shearing and non-rotating. In all these models, we observe that they do 
not approach isotropy for large values of time $t$ either in the presence or in the 
absence of magnetic field. \\
The cosmological constant in all models given in Sec. 4 are decreasing 
function of time and they all approach a small value as time increases (i.e., the
present epoch). The values of cosmological ``constant'' for these models are
found to be small and positive which are supported by the results from recent
supernovae observations. If we consider $\Lambda$ as constant and coefficient
of bulk viscosity $\xi$  as zero, we obtain the solutions obtained by Bali and
Jain.\cite{ref62} If we set $\Lambda = constant$, $\xi = 0$ and $n = 1$, we get the
solutions obtained by Bali.\cite{ref71} Thus, with our approach, we obtain a 
physically relevant decay law for the cosmological constant unlike other investigators 
where {\it adhoc} laws were used to arrive at a mathematical expressions for 
the decaying vacuum energy. Thus our models are more general than those studied earlier.\\ 
\nonumsection{Acknowledgements}
\noindent A. Pradhan thanks to the Inter-University Centre for Astronomy and Astrophysics, 
India for providing facility under Associateship Programme where part this work was 
carried out. Authors would like to thank R. G. Vishwakarma for helpful discussions.\\
\newline
\newline
\nonumsection{References}

\end{document}